# Optical anisotropy induced by torsion stresses in LiNbO$_3$ crystals: appearance of an optical vortex

Ihor Skab, Yurij Vasylkiv, Viktoriya Savaryn and Rostyslav Vlokh[*]

*Institute of Physical Optics, 23 Dragomanov St., 79005 Lviv, Ukraine*

[*]*Corresponding author: vlokh@ifo.lviv.ua*

We report the results of studies of torsion effect on the optical birefringence in LiNbO$_3$ crystals. We have found that twisting of those crystals causes a birefringence distribution revealing non-trivial peculiarities. In particular, it has a special point at the center of cross section perpendicular to the torsion axis where zero birefringence value occurs. It has also been ascertained that the surface of the spatial birefringence distribution has a conical shape, with the cone axis coinciding with the torsion axis. We have revealed that an optical vortex, with the topological charge equal to unity, appears under the torsion of LiNbO$_3$ crystals. It has been shown that, contrary to the *q*-plate, both the efficiency of spin-orbital coupling and the orbital momentum of the emergent light can be operated by the torque moment.

# 1. Introduction

The optical vortices can usually be induced using different experimental techniques and media, in particular computer-generated holograms, optical fibers, etc [1-4]. In the most of cases the media mentioned above are optically isotropic and inhomogeneous, i.e. they reveal some distribution of optical density. Then an initially nearly plane wave undergoes a perturbation of its phase front, with appearance of scalar field singularities or optical vortices during propagation through these media. Another kind of singularities, singularities of vector fields, can be observed when different types of beams, such as paraxial Gaussian, Laguerre–Gaussian or Bessel–Gaussian beams, propagate in optically uniaxial anisotropic media [5–10], or in biaxial crystals provided that the beam propagates along the optic axis under the conditions of conical refraction [11–13].

In the case of uniaxial crystals, the polarization structure of a circularly polarized Gaussian beam propagating along the optic axis contains a degenerate *C*-point (a point with a circular polarization [14]), with a topological index $p = \pm 1$, which is associated with a doubly charged optical vortex in one of the circularly polarized components [8, 15]. Any inclination of the beam axis with respect to the optic axis results in slipping the polarization singularity off the axis and leaving the beam. Besides, the energy efficiency of vortex generation in the beam traveling along the optic axis depends strongly on the beam waist radius [16]: this efficiency tends quickly to zero with increasing beam radius. In the limiting case, we come to an obvious conclusion: a plane wave propagating along the optic axis cannot gain any polarization singularity.

However recently it has been shown [17] that it is possible to convert a spin momentum carried by a circularly polarized light beam into orbital angular momentum with the aid of a liquid crystal matrix (a so-called *q*-plate), thus leading to generation of helical modes with a sign of wave-front helicity controlled by the sign of input circular polarization. In other words, the input polarization of light controls the sign of the orbital helicity of the output wave front. Its magnitude is instead fixed by the birefringence axis geometry. This phenomenon requires interaction of light with a matter that is both optically inhomogeneous and anisotropic. Obviously, the method of *q*-plates does not allow controlling the angular momentum of the emergent light. Then a following question arises: can a polarization singularity be embedded into the plane wave itself (i.e., a very wide Gaussian beam) provided that the uniaxial crystal is subjected to inhomogeneous action inducing an axially symmetric gradual distribution of, e.g., refractive index? Or, in somewhat different terms, is it possible to create a vortex in the initially nearly plane wave using anisotropic inhomogeneous media under external actions, in order to control the angular momentum of the emergent beam? It is evident that mechanical torsion around the optic axis (not to be confused with a rotation of the axes of anisotropy earlier studied in Refs. [18, 19]) belongs to the actions which can create gradual distribution of, e.g., the refractive index [20]. Then one would control the angular momentum of the emergent beam by controlling the torque moment.

Following from the constitutive equation $D_i = B_{ij}E_j$ describing propagation of light in anisotropic media (with $E_j$ being the electric field and $D_i$ the electric displacement of an electromagnetic wave propagating in a transparent, magnetically non-ordered medium), which reveal no imaginary part of $B_{ij}$ caused by absorption or magnetic moments, one can derive a well known relation for a piezooptic effect. The latter consists in changes $\Delta B_{ij}$ of the optical impermeability coefficients (or the refractive indices, since $B_{ij} = (1/n^2)_{ij}$) imposed by the action of mechanical stresses $\sigma_{kl}$ [21]:

$$\Delta B_{ij} = B_{ij} - B_{ij}^0 = p_{ijkl}\sigma_{kl}, \qquad (1)$$

where $p_{ijkl}$ is a fourth-rank piezooptic tensor, and $B_{ij}$ and $B_{ij}^0$ the impermeability tensors of a stressed and free samples, respectively.

Usually the piezooptic effect is investigated under homogeneous stresses applied to samples. As we have shown in our works [20, 22–26], application of torsion [20, 22, 24, 25] or bending [20, 23, 25] stresses can lead to a number of new nontrivial crystal optical effects, which are induced by non-zero coordinate derivatives of those stresses, $\partial \sigma_{kl}/\partial x_m$, rather than by the uniform stresses themselves. One of these effects, a gradient piezogyration (or a torsion-gyration effect), has been found in NaBi(MoO$_4$)$_2$ crystals [20, 26]. Contrary to a common piezogyration (see [27]), this effect is caused only by inhomogeneous stresses. It leads to appearance of optical activity in crystals under torsion stresses and may be described by the relation $\Delta g_{ij} = g_{ij} - g_{ij}^0 = \Theta_{ijklm} \partial s_{kl} / \partial x_m$, where $\Delta g_{ij}$ is an increment of the gyration tensor with respect to its initial value, $g_{ij}^0$, and $\Theta_{ijklm}$ denotes a fifth-rank axial tensor.

Spatial distributions of the optical birefringence and the optical indicatrix rotation in LiNbO$_3$ crystals have also been studied in part for a particular case of laser beam propagating along the optic axis and the torsion applied around the same direction [23, 28]. What is the most important, a special point of zero birefringence belonging to the torsion axis has been revealed in the geometrical center of *XY* cross section of a sample, which corresponds to zero shear stress components $s_{13}$ and $s_{23}$. The birefringence has been found to increase with increasing distance from that geometrical center and so it has been supposed that the birefringence distribution has a conical shape in the coordinates $X, Y, \Delta n$ [20, 22].

It has also been established that the induced birefringence changes its sign while crossing the origin of the *XY* cross section along any diameter. Moreover, a character of spatial distribution of the optical indicatrix rotation angles observed in the experiments [20, 22] suggests that, at least at the points defined by the coordinates (*X*, 0), (–*X*, 0) and (0, *Y*), (0, –*Y*), the optical indicatrices are rotated by $90\deg$. In other words, the indicatrices have mutually orthogonal orientations on diametrically opposite sides with respect to the origin of the coordinate *XY* system, thus meaning that the sign of the birefringence should inevitably change at the origin. The latter facts contradict a simple conical distribution of the birefringence that might have been supposed. Besides, the *XY* distribution of the optical indicatrix orientation for the LiNbO$_3$ crystal twisted around the *Z* axis seems to be not completely comprehensible, since it has some regions where the optical indicatrices change their orientation drastically, with the orientation angles jumping, e.g., from $-45$ to $+45\,\text{deg}$. Such a spatial distribution of the optical indicatrix orientation appears to be at evident variance with the condition of continuous medium, at least at the boundaries among the regions with the orthogonal orientations.

It is worthwhile to remember that a considerable drawback of our previous experiments [20, 22] has been utilization of a single laser beam polarimetric technique, when the beam is being successively scanned across the *XY* face of a sample. This technique manifests a low resolution limited by a laser beam diameter, is quite routine and, moreover, the measurement procedures are durable and difficult for automation. At the same time, very interesting are the facts that the torsion imposes a peculiar distribution of birefringence characterized with a singular point $\Delta n = 0$ in the center of sample cross section and an increase occurring when moving away from the torsion axis. The latter can be used for creation of a singular beam bearing optical vortex under the condition of crystal torsion.

## 2. Experimental methods

A sample of LiNbO$_3$ crystal (the point symmetry group 3*m*) used in our experiment was prepared as an octahedral prism, with its lateral faces parallel to *Z* axis (see Fig. 1) and the basis parallel to *XY* plane. The sample had the sizes of 13 mm along the optic axis *Z* and 6 mm between the lateral faces. The *YZ* plane was accepted to be parallel to one of the symmetry mirror planes. A light of He-Ne laser (the light wavelength of $l = 632.8$ nm) propagated along the *Z* axis.

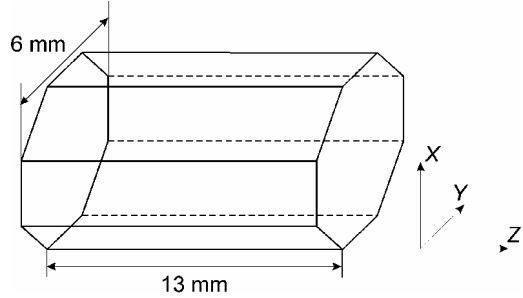

Fig. 1. Shape and orientation of LiNbO$_3$ crystal sample.

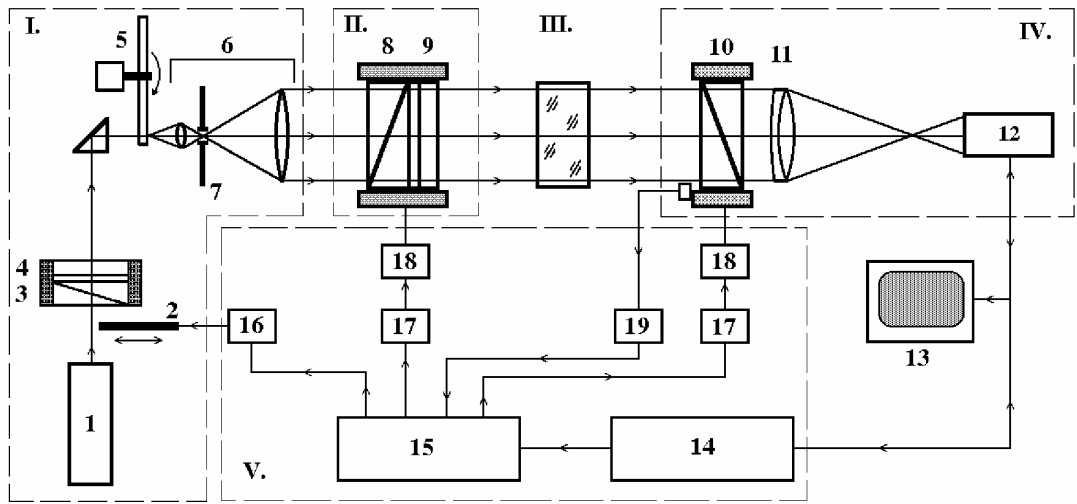

Fig. 2. Schematic representation of our imaging polarimeter (I – light source section; II – polarization generator, III – sample section, IV – polarization analyzer, and V – controlling unit): 1 – He-Ne laser; 2 – ray shutter; 3, 8 – polarizers; 4, 9 – quarter-

wave plates; 5 – coherence scrambler; 6 – beam expander; 7 – spatial filter; 10 – analyzer; 11 – objective lens; 12 – CCD camera; 13 – TV monitor; 14 – frame grabber; 15 – PC; 16 – shutter's controller; 17 – step motors' controllers; 18 – step motors; 19 – reference position controller.

Torsion torques $M_z$ were applied to one of the basic faces of the crystalline prism, while the opposite basic face was kept to be fixed. Our polarimetric experimental setup is shown in Fig. 2. We used a number of different experimental techniques basing on the same apparatus: a method for determining spatial distribution of the optical indicatrix orientation, two methods for the birefringence distribution studies, and a method for revealing the optical vortices.

While studying the optical indicatrix rotation, we exploited an imaging polarimeter described earlier (see [29] and Fig. 2), with the following minor modifications: we removed a quarter-wave plate 9 (see Fig. 2) from the polarization generator and used a linearly polarized incident light. An analyzer 10 was placed into the extinction position with respect to a polarizer. The crossed polarizers were simultaneously rotated in the range of 180 deg with the step 5 deg. The angles corresponding to the minimums of transmitted light intensity detected by a CCD camera were ascribed to the extinction positions or the orientations of optical indicatrices in each part of sample image.

One of the techniques used for measuring the birefringence was a well-known Senarmont method [30]. An additional, when compare with Fig. 2, quarter-wave plate 9 was then placed between a sample section III and analyzer 10. The birefringence was calculated with the formula $\Delta n = \beta \lambda / \pi d$, where $\boldsymbol{b} = \boldsymbol{D\Gamma}/2$ means the rotation angle of polarization plane for the light transmitted through the quarter-wave plate with respect to the corresponding initial angle, $\boldsymbol{D\Gamma}$ the phase difference, and $d$ the sample thickness along the direction of light propagation.

When using the second technique, we made the probing beam circularly polarized, since it is not sensitive to the orientation of optical indicatrix. Then the quarter-wave plate 9 was remained in its initial position as shown in Fig. 2. The angle between the principal axes of the quarter-wave plate and the transmission direction of polarizer 8 was equal to 45 deg. In this case the sample is described by a model of linear phase retarder for which the dependence of the output intensity $I$ on the analyzer azimuth $\boldsymbol{a}$ is expressed as

$$I = \frac{I_0}{2}\{1 + \sin \boldsymbol{D\Gamma} \sin[2(\boldsymbol{a} - \boldsymbol{\varphi})]\} = C_1 + C_2 \sin[2(\boldsymbol{a} - C_3)], \qquad (2)$$

where $\boldsymbol{\varphi}$ is the orientation angle of the optical indicatrix and $\boldsymbol{D\Gamma} = 2\pi \boldsymbol{D}n d / \lambda$ the optical phase retardation. After recording and filtering image, azimuthal dependences of the intensity $I$ were fitted by the sine function for every pixel of the image, with the fitting coefficients

$$C_1 = \frac{I_0}{2}, \ C_2 = \frac{I_0}{2}\sin \Delta\Gamma, \ C_3 = \varphi. \qquad (3)$$

Now it is seen that the optical retardation $D\Gamma$ is given by the coefficients $C_1$ and $C_2$:

$$\sin D\Gamma = C_2 / C_1, \qquad (4)$$

while the angular orientation of the intensity minimum is determined by the orientation of principal axis φ of the optical indicatrix and the coefficient $C_3$. Hence, fitting of dependences of the light intensity after the analyzer upon the azimuth for each pixel of the sample image enables constructing 2D maps of the optical anisotropy parameters of the sample, the optical retardation and the orientation of the optical indicatrix.

Finally, when checking the appearance of optical vortices, we used a wide circularly polarized incident beam propagating along the Z axis of LiNbO$_3$ crystals subjected to torsion around the same direction. A circular polarizer was used behind a crystalline sample, with its circularity sign opposite to that of the input polarizer. Then an additional quarter-wave plate was placed between a sample section III and analyzer 10.

## 3. Experimental results

The map of the optical indicatrix rotation in the LiNbO$_3$ crystals observed under the torsion torque $M_z = 63.77 \times 10^{-3}$ N×m is depicted in Fig. 3a. The optical indicatrix rotates around the torsion axis by the angles ranging from 0 deg in the $Y = 0$ ($Y > 0$) plane ($\varphi = 0 \deg$) to 45 deg at $X = 0$ ($X > 0$) plane ($\varphi = 90 \deg$), with $\varphi$ denoting the angle defined in the polar coordinate system ($r, \varphi$) via $X = r\cos\varphi$ and $Y = r\sin\varphi$. In other words, the optical indicatrix rotates around the torsion axis by the angles $\zeta_z = \varphi/2$. The indicatrices arranged at the diameters of the *XY* cross section and located on the opposite sides with respect to the torsion axis (the origin of the *XY* coordinate system) are rotated by 90 deg. Besides, the sign of the optical indicatrix rotation depends on the sign of the torsion torque (see Fig. 3b and Fig. 3c), i.e. application of the torsion torques $M_z$ and $-M_z$ (with the sign "plus" being attributed to the torque inducing a right-handed screw) yields the opposite rotations.

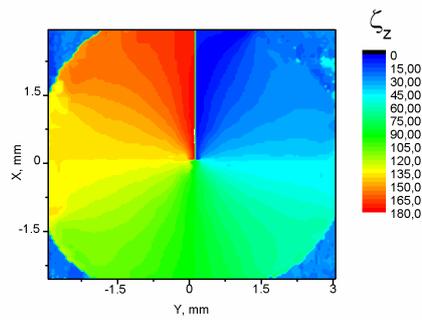

(a)

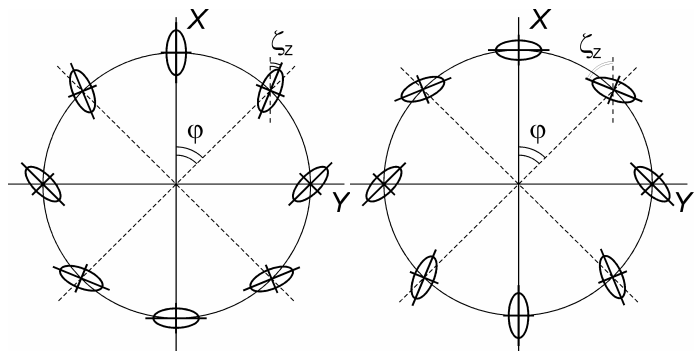

(b)           (c)

Fig. 3. Spatial distributions of optical indicatix rotation in LiNbO$_3$ crystals at $l = 632.8$ nm: (a) the pattern observed experimentally under the torsion torque $M_z = 63.77 \times 10^{-3}$ N×m, (b) and (c) the patterns appearing for the two opposite torsion torques which are simulated using Eqs. (11) and (13). Orientations of optical indicatrices are denoted by ellipses.

The spatial distributions of the birefringence induced by the torsion torque $M_z = 63.77 \times 10^{-3}$ N×m applied along the *X*, *Y* axes and the bisector of those axes, which have been obtained with the Senarmont technique, are presented in Fig. 4. As an example, in Fig. 4b we show also a distribution of the rotation angle of the light polarization plane behind the quarter-wave plate obtained after scanning along the *X* axis. These dependences are exactly linear, at least for the distances less than ~2 mm from the torsion axis. At larger distances, deviations from a linear dependence are observed. They can be readily explained by the influence of sample boundaries and deviation of the sample shape from cylindrical one.

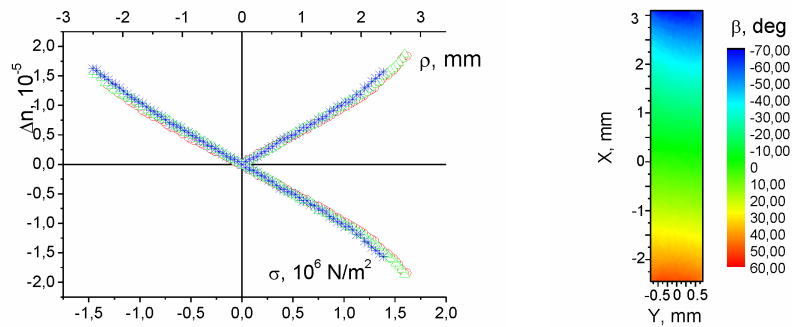

(a) (b)

Fig. 4. (a) Spatial distribution of birefringence induced by the torsion torque $M_z = 63.77 \times 10^{-3}$ N×m (open circles correspond to scanning along the X axis under the action of $s_{32}$ stress, open triangles – scanning along the Y axis under the action of $s_{31}$ stress, and crosses – scanning along the bisector of the X and Y axes under the action of $s = s_{31} = s_{32}$ stress; a scale corresponding to the shear stress components is also shown; dependence in the upper right side corresponds to the induced birefringence rewritten in the coordinate system associated with eigenvectors of the optical indicatrix) and (b) distribution of rotation angle (in angular deg) of the light polarization plane behind the quarter-wave plate in case of scanning along the X axis.

On the contrary, a 2D distribution of the birefringence obtained with the second experimental technique (see Fig. 5) manifests no change in the sign when crossing the geometrical center of the XY cross section. The contradiction of results shown in Fig. 4 and Fig. 5 may be easily explained as follows. When obtaining the coordinate dependences of the birefringence with the Senarmont technique (Fig. 4), we have associated both the polarizer and quarter-wave plate with the laboratory coordinate system rather than the eigenvectors of the Fresnel ellipsoid. At the same time, the optical indicatrix is rotated by 90 deg while crossing the origin of the XY coordinate system, corresponding to interchange of the eigen axes X and Y. That is why no change in the birefringence sign is seen if we rewrite the induced birefringence in the coordinate system associated with the eigenvectors of the optical indicatrix, in agreement with the results depicted in Fig. 5. On the other hand, in the conditions of our experiments we can neglect deviations in the optical beam that could appear due to inhomogeneous radial distribution of the refractive indices. This is because the lateral shift of the beams emergent from the sample is smaller than the size of pixel of the CCD camera, being of the order of $10^{-6}$ m.

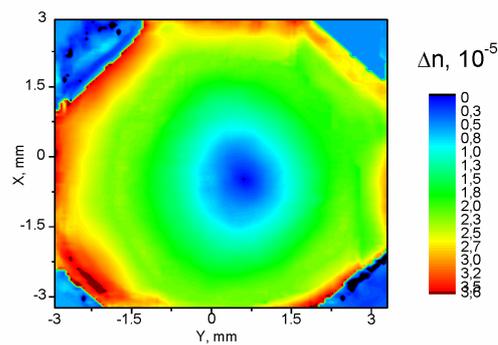

Fig. 5. Experimental distribution of birefringence induced by the torsion torque $M_z = 63.77 \times 10^{-3}$ N×m in the *XY* plane of LiNbO$_3$ crystals at 632.8 nm.

## 4. Discussion

### *4.1. Orientation of optical indicatrix*

Under the torsion of cylindrical sample, the stress tensor components may be determined as [31]

$$\sigma_m = \frac{2M_z}{\pi R^4}(X d_{4m} - Y d_{5m}), \quad (5)$$

where a difference in the stress distributions occurred for cylindrical and octahedral samples is neglected. Here $M_z = \int_S r \times P dS$, $d_{nm}$ is the Kronecker delta, $R$ the cylinder radius, $S$ the square of the cylinder basis, and $P$ is the mechanical load. Therefore we have two shear components of the stress tensor, $\sigma_4 = \sigma_{32}$ and $\sigma_5 = \sigma_{31}$:

$$\sigma_{32} = \frac{2M_z}{\pi R^4} X, \quad (6)$$

$$\sigma_{31} = \frac{2M_z}{\pi R^4} Y. \quad (7)$$

which depend linearly on the coordinates.

The piezooptic tensor for the point group 3m is as follows:

|            | $\sigma_{11}$ | $\sigma_{22}$ | $\sigma_{33}$ | $\sigma_{32}$ | $\sigma_{31}$ | $\sigma_{21}$ |
|------------|---------------|---------------|---------------|---------------|---------------|---------------|
| $\Delta B_{11}$ | $\pi_{11}$ | $\pi_{12}$ | $\pi_{13}$ | $\pi_{14}$ | 0 | 0 |
| $\Delta B_{22}$ | $\pi_{11}$ | $\pi_{11}$ | $\pi_{13}$ | $-\pi_{14}$ | 0 | 0 |
| $\Delta B_{33}$ | $\pi_{31}$ | $\pi_{31}$ | $\pi_{33}$ | 0 | 0 | 0 |
| $\Delta B_{32}$ | $\pi_{41}$ | $-\pi_{41}$ | 0 | $\pi_{44}$ | 0 | 0 |
| $\Delta B_{31}$ | 0 | 0 | 0 | 0 | $\pi_{44}$ | $2\pi_{41}$ |
| $\Delta B_{21}$ | 0 | 0 | 0 | 0 | $\pi_{14}$ | $\pi_{66}$ |

(8)

Then the equation of optical indicatrix perturbed by the two shear stresses may be written as

$$(B_{11} + \pi_{14}\sigma_{32})X^2 + (B_{11} - \pi_{14}\sigma_{32})Y^2 + B_{33}Z^2$$
$$+ 2\pi_{44}\sigma_{32}YZ + 2\pi_{44}\sigma_{31}XZ + 2\pi_{14}\sigma_{31}XY = 1 \quad . \quad (9)$$

The angle of optical indicatrix rotation around the Z axis is given by

$$\tan 2V_z = \frac{s_{31}}{s_{32}}. \tag{10}$$

In the particular case of $X = 0$ and $s_{32} = 0$ one has

$$\tan 2V_z = \frac{2p_{14}s_{31}}{B_{11} - B_{11}} = \frac{2p_{14}}{B_{11} - B_{11}} \frac{2M_z}{pR^4} Y \to \pm\infty, \quad V_z \to \pm 45°, \tag{11}$$

where the signs '$\pm$' correspond to different torque sign, while under the conditions $Y = 0$ and $s_{31} = 0$ the relations

$$\tan 2z_z = 0, \quad z_z = 0 \tag{12}$$

hold true. Taking formulae (6) and (7) and definitions $X = r\cos j$, $Y = r\sin j$ into account, we rewrite formula (10) as

$$\tan 2V_z = \frac{s_{31}}{s_{32}} = \frac{Y}{X} = \frac{\sin j}{\cos j} = \tan j . \tag{13}$$

Thus, we arrive at a simple dependence $z_z = j/2$. As seen from Fig. 3a, the experimental results exactly satisfy this relation. Actually, in the case of the clockwise path tracing on the angle $j$, beginning from the positive X axis direction, the optical indicatrix rotates by the angle equal to $j/2$ (see Fig. 3).

## *4.2. Spatial distribution of birefringence*

In order to analyze spatial distribution of the birefringence in the XY plane, it is necessary to consider a cross section, by the plane $Z = 0$, of the Fresnel ellipsoid given by formula (9):

$$(B_{11} + p_{14}s_{32})X^2 + (B_{11} - p_{14}s_{32})Y^2 + 2p_{14}s_{31}XY = 1. \tag{14}$$

Thence the relations for the principle refractive indices strictly follow:

$$n_1 = n_o + \frac{1}{2}n_0^3 p_{14}\sqrt{s_{31}^2 + s_{32}^2},$$
$$n_2 = n_o - \frac{1}{2}n_0^3 p_{14}\sqrt{s_{31}^2 + s_{32}^2}. \tag{15}$$

where $n_o$ is the initial (stress-free sample) ordinary refractive index. Remembering formulae (6) and (7), we can rewrite the refractive indices as functions of the Cartesian coordinates,

$$n_1 = n_0 + n_0^3 p_{14}\frac{M_z}{pR^4}\sqrt{Y^2 + X^2},$$
$$n_2 = n_0 - n_0^3 p_{14}\frac{M_z}{pR^4}\sqrt{Y^2 + X^2}, \tag{16}$$

and the polar coordinates,

$$n_1 = n_0 + n_0^3 p_{14}\frac{M_z}{pR^4}r,$$
$$n_2 = n_0 - n_0^3 p_{14}\frac{M_z}{pR^4}r. \tag{17}$$

Then the birefringence can be obtained:

$$\Delta n = n_0^3 p_{14}\sqrt{s_{31}^2 + s_{32}^2} = 2n_0^3 p_{14}\frac{M_z}{pR^4}\sqrt{Y^2 + X^2} = 2n_0^3 p_{14}\frac{M_z}{pR^4}r. \tag{18}$$

As a result, the birefringence is not a function of the $j$ angle, though it depends on the module $r$, i.e. the distance from the torsion axis (see Fig. 5 and Fig. 6). Again, the latter fact agrees perfectly with the spatial birefringence distribution presented in Fig. 4 and Fig. 5. Besides, formula (18) just represents an equation of a cone with its apex located at the origin of the coordinate system (Fig. 6). Thus, according to the experimental results (Fig. 5 and Fig.6), the path tracing involving variations of only the angle $j$ causes no birefringence sign change and, subsequently, no changes in the phase of light. In fact, the only point of the map characterized by a singular birefringence value is a geometrical center of the sample cross section, which is equal to zero. On the other hand, distribution of the emergent light polarization should possess a vector field singularity under the conditions $X = Y = 0$.

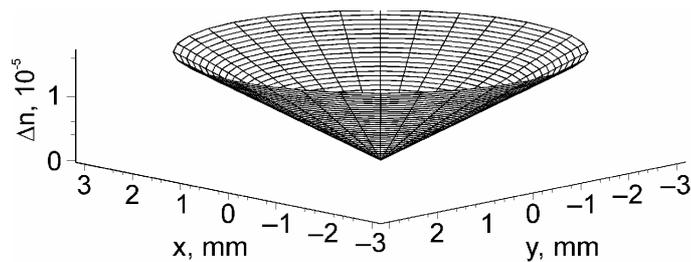

Fig. 6. Experimental conical spatial distribution of optical birefringence appearing under the torsion of LiNbO$_3$ crystals (the torsion torque $M_z = 63.77 \times 10^{-3}$ N×m and $l$ =632.8 nm ).

Distribution of the birefringence on the coordinates $X$ and $Y$ should manifest itself in a double cone-like distribution of the refractive indices. Really, the dependence of the refractive indices on the coordinates $X$ and $Y$ should form a double conical surface, with the common apex at the point $X = Y = 0$. This behavior is somewhat similar to the effect of conical refraction. However, in the latter case the refractive indices are not functions of the coordinates for a homogeneous crystal, while the double conical distribution mentioned above is associated with the angular coordinates of the light wave vector. In the present situation such a distribution is generated if a paraxial beam is dealt with, owing to dependence of the birefringence on the coordinates *X* and *Y*, along with equality of the latter to zero at the origin (a so-called "diabolical point" – see [11]). Notice also that in the case of the conical refraction a vortex with the unit charge appears if the circular polarized beam is incident [12].

### *4.3. Observation of vortex appearing under torsion of LiNbO$_3$ crystals*

In order to examine the assumption whether the LiNbO$_3$ crystals subjected to the torsion stress around the *Z* axis can generate a vortex in the initially parallel beam, we have used the next sequence of optical elements: a right-handed circular polarizer, a sample under torsion, and a left-handed circular analyzer. It is evident from Fig. 7 that the light intensity in the center of the image remains zero, irrespective of the torsion torque, a bright ring around a central dark circle appears and becomes more distinguishable with increasing torque, while the diameter of the central dark region decreases. Such a behavior corresponds to a vortex carried by the emergent light.

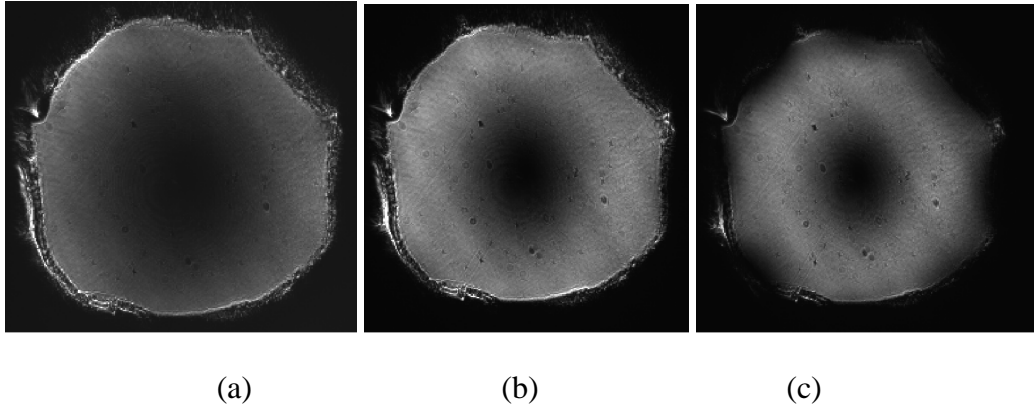

(a)            (b)            (c)

Fig. 7. CCD image of expanded beam emergent from the system consisting of a right-handed circular polarizer, a twisted sample and a left-handed circular polarizer (a – $M_z$=63.77×10$^{-3}$ N×m, b – $M_z$=98×10$^{-3}$ N×m, and c – $M_z$=162×10$^{-3}$ N×m).

The contrast between the central dark ring and the bright ring increases with increasing torque moment. At the same time, the central dark spot becomes narrow. Such a behavior suggests that the increase in the torque moment is accompanied by decreasing total spin angular momentum defined by the relation $S_z = (I_r - I_l)/(I_r + I_l)$ [32] (with $I_r$ and $I_l$ standing for the total intensities of the right-hand and left-hand polarized beam components, respectively). Hence, the orbital angular momentum of the emergent beam should increase with increasing torque moment, owing to the momentum conservation law.

    Let us now analyze distribution of the optical retardation induced by the torsion torque, which corresponds to the results depicted in Fig. 7. Simulations of this distribution have been carried out using the Jones matrix approach. The sample cross section under study has been divided by 3600 (60×60) elementary, optically

uniform cells. The resulting Jones matrices for each of the elementary beams ($k,l=1...60$) have been obtained as follows:

$$J_{kl} = \begin{vmatrix} \left(e^{i\Delta\Gamma_{kl}/2}\cos^2 x_{kl} + e^{-i\Delta\Gamma_{kl}/2}\sin^2 x_{kl}\right) & i\sin(\Delta\Gamma_{kl}/2)\sin 2x_{kl} \\ i\sin(\Delta\Gamma_{kl}/2)\sin 2x_{kl} & \left(e^{i\Delta\Gamma_{kl}/2}\sin^2 x_{kl} + e^{-i\Delta\Gamma_{kl}/2}\cos^2 x_{kl}\right) \end{vmatrix}, \quad (19)$$

$$(s_{32})_{kl} = \frac{2M_z}{pR^4}\left(\frac{k-30}{10}10^{-3}\right), \quad (s_{31})_{kl} = \frac{2M_z}{pR^4}\left(\frac{l-30}{10}10^{-3}\right),$$
$$DG_{kl} = 2pd\left\{n_0^3 p_{14}\sqrt{(s_{32})_{kl}^2 + (s_{31})_{kl}^2}\right\}\Big/l, \quad (20)$$
$$x_{kl} = \frac{1}{2}\arctan\frac{(s_{32})_{kl}}{(s_{31})_{kl}},$$

where $M_z = 63.77 \times 10^{-3}$ N×m, $n_o = 2.28647$ [33], $R = 3$ mm, $d = 13$ mm and $|p_{14}| = (8.87 \pm 0.28) \times 10^{-13}$ m$^2$/N [34].

Let $E_1$, $E_2$ and $E_1^{kl}$, $E_2^{kl}$ be the components of the input and output Jones vectors, respectively, and $J^{QWP-}$, $J^{QWP+}$ and $J^A$ be the Jones matrices of the left handed circular polarizer, right handed circular polarizer and analyzer, respectively. Then we have:

$$\begin{vmatrix} E_1^{kl} \\ E_2^{kl} \end{vmatrix} = J^A J^{QWP-} J^{kl} J^{QWP+} \begin{vmatrix} E_1 \\ E_2 \end{vmatrix}, \quad (21)$$

where

$$E_1 = 1, \quad E_2 = 0, \quad J^A = \begin{pmatrix} 0 & 0 \\ 0 & 1 \end{pmatrix},$$

$$J^{QWP-} = \begin{pmatrix} \frac{1}{\sqrt{2}} e^{i\frac{\pi}{4}} & \frac{1}{\sqrt{2}} e^{-i\frac{\pi}{4}} \\ \frac{1}{\sqrt{2}} e^{-i\frac{\pi}{4}} & \frac{1}{\sqrt{2}} e^{i\frac{\pi}{4}} \end{pmatrix}, \quad J^{QWP+} = \begin{pmatrix} \frac{1}{\sqrt{2}} e^{-i\frac{\pi}{4}} & \frac{1}{\sqrt{2}} e^{i\frac{\pi}{4}} \\ \frac{1}{\sqrt{2}} e^{i\frac{\pi}{4}} & \frac{1}{\sqrt{2}} e^{-i\frac{\pi}{4}} \end{pmatrix},$$

(22)

the resulting optical retardation for each of the elementary beams may be determined by the relation

$$\boldsymbol{DG}^{kl} = \arctan\left(\frac{\operatorname{Im} E_1^{kl}}{\operatorname{Re} E_1^{kl}}\right) - \arctan\left(\frac{\operatorname{Im} E_2^{kl}}{\operatorname{Re} E_2^{kl}}\right).$$

(23)

The spatial distribution of the optical retardation (phase difference) simulated with formula (23) is presented in Fig. 8. It is seen from Fig. 8 that the path tracing along the angle $\varphi$ around the geometrical center of this image changes the phase difference (or the phase of optical wave) by the same angle [35-37]. Thus the phase difference, which is indefinite at the center of the map may be represented as a helicoid. It is interesting to notice that, since the phase difference angle coincides with the tracing angle, the vortex induced by the torsion in our case has its topological charge equal to unity. It is obvious that the sign of the vortex charge would depend on the signs of both the torsion torque and the incident circularly polarized wave. In other words, the sign of the charge would be positive if the circular wave and the torsion torque have the same signs, and negative otherwise. Indeed, the vortex described above is analogical to those appearing in the case of conical refraction [12, 13, 38]. However, in our case it is caused by inhomogeneous spatial distribution of the birefringence.

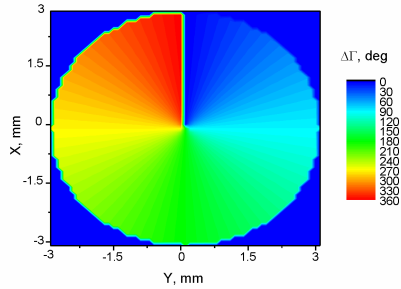

Fig. 8. Simulated phase difference distribution induced by the torsion torque $M_z = 63.77 \times 10^{-3}$ N×m in the $XY$ plane of LiNbO$_3$ crystals at 632.8 nm.

## Conclusions

When studying the effect of torsion on the birefringence and optical indicatrix rotation in LiNbO$_3$ crystals, we have found that the resulting spatial distribution of the birefringence reveals non-trivial peculiarities. Namely, it has a special point with zero birefringence at the center of the sample cross section perpendicular to the torsion axis. In other words, the birefringence is not induced by the torsion torque at the very torsion axis. The magnitude of the birefringence increases linearly while moving out from the center of the cross section. This is in agreement with the solution of equation for the perturbed optical indicatrix (see Eqs. (9) and (18)). The contradiction concerned with the change in birefringence sign measured by the Senarmont technique while crossing geometrical center of the $XY$ cross section of LiNbO$_3$ crystals and conical distribution of the birefringence is solved in the present work. It is demonstrated that really there is no contradiction between the latter facts since we have associated both the polarizer and the quarter-wave plate with the laboratory coordinate system rather than the eigenvectors of the Fresnel ellipsoid when obtaining the coordinate dependences of the birefringence with the Senarmont technique. Thus, the spatial distribution of the birefringence indeed forms a conical surface.

We have shown that the spatial surface of the birefringence distribution built in the coordinates ($X, Y, \Delta n$) has a conical shape, where the cone axis is given by the torsion direction. We have also found experimentally that an optical vortex with the topological charge equal to unity appears in the optical system consisting of a right-handed circular polarizer, a LiNbO$_3$ crystal twisted around the $Z$ axis and a left-handed circular polarizer. It has been shown that, contrary to the $q$-plate, both the efficiency of spin-orbital coupling and the orbital momentum of the emergent light can be operated by the torque moment.

## Acknowledgement

We thank Prof. A. Volyar for useful discussion.